# *In situ* soft x-ray absorption spectroscopic study of Fe/MgO interfaces

Pramod Vishwakarma[1], Mukul Gupta[2], D.M. Phase[2] and Ajay Gupta[1,*]

[1] *Amity Centre for Spintronic Materials, Amity University UP, Sector 125, Noida 201 313*

[2] *UGC-DAE Consortium for Scientific Research, University Campus, Khandwa Road, Indore 452 001*

*Corresponding author: e-mail: agupta2@amity.edu

**ABSTRACT**

Interfacial structure between iron and MgO has been studied *in situ* during deposition of iron on MgO surface, using soft x-ray absorption spectroscopy. Submonolayer sensitivity of the technique combined with the *in situ* measurements as a function of iron layer thickness allowed one to follow evolution of interfacial region. Two different substrates namely, MgO (001) single crystal, and a polycrystalline MgO film on Si substrate have been used in order to elucidate the role of the state of MgO surface in controlling the interface structure. It is found that at the interface of iron and MgO film, interfacial interaction results in formation of $Fe_3O_4$, while on MgO (001) surface iron mono-oxide is formed. Thickness of the interfacial layer has been determined with submonolayer accuracy. $Fe_3O_4$ being the oxide of iron with the highest heat of formation, the evolution of the interface of Fe on MgO film appears to be controlled thermodynamically. On the other hand, on MgO (001) surface, interfacial reaction is limited by the availability of oxygen atoms. A comparison with earlier results suggest that magnetic behaviour of the FeO layer gets modified significantly due to proximity effect of the bulk ferromagnetic iron layer.

# 1. INTRODUCTION

Multilayer nanostructures involving various metals and oxides are important materials for a variety of applications in microelectronics, spintronics, and as catalysts, sensors etc. [1,2]. Interfacial interactions between metals and oxide thin films play a key role in controlling their functional properties [3,4]. Therefore, there is a great interest in understanding such interfacial interactions and their correlation with the functional properties. Fe/MgO system has been a subject of extensive investigations for last more than a decade, as a number of interesting phenomena have been observed in this system: (i) Firstly, it was shown theoretically and subsequently confirmed experimentally that because of coherent tunneling effect at the interface between Fe and MgO a tunnel magnetoresistance several orders of magnitude higher than that expected on the basis of Jullière's model can be achieved [5-7]. (ii) Subsequently, it was found that ultra-thin film of Fe on MgO possesses a perpendicular magnetic anisotropy (PMA) [2,8-10]. This gave rise to the possibility of preparing perpendicular magnetic tunnel junctions (MTJ) which are superior to the in-plane magnetic tunnel junction in terms of both thermal stability and the scalability to smaller dimensions, as well as in terms of more energy efficient way of switching the magnetization direction [2,11,12]. (iii) More recently, demonstration of the possibility of voltage control of PMA in Fe/MgO system [13-16] has led to the possibility of low power manipulation of magnetization. In recent years a few more interesting effects have been observed in this system, for example, interlayer coupling between Fe layers through MgO [17-19] and the exchange bias effect due to possible interfacial oxidation [20].

It may be noted that all the above effects are highly sensitive to the structure of the interface between Fe and MgO layers. A possible interdiffusion or oxidation of iron even over a few monolayers can drastically alter the behaviour of the system. For example, it has been shown theoretically that formation of iron oxide or structural defects/vacancies occurring over even 1 or 2 monolayers at the interface can drastically affect the coherent tunneling of electrons across the interface [21-23]. The PMA too has its origin in the anisotropic hybridization of Fe with O at the interface [2,24]. Formation of some oxide of magnetic species at the interface either during the deposition process or intentionally done post-deposition, [25,26] can drastically affect the PMA. The fact that PMA has its origin in interfacial hybridization also lies in the root of its control through an applied voltage [13,15,27]. Theoretical calculations show that oxidation of Fe at Fe/MgO interface plays an important role in electric field control of interfacial magnetic anisotropy; [28,29] Even the oxidation state of magnetic ions at the interface with MgO can be controlled in a reversible manner using electric field [26]. The observed exchange bias of Fe moments in the interfacial region has also been attributed to possible formation of antiferromagnetic oxide of Fe at the interface [20]. Therefore, extensive studies are being done in order to characterize the interface of Fe/MgO with high accuracy. However highly contradictory result have been reported in the literature; While several studies suggest that there is no oxidation of iron or formation of any magnetically dead layer at the interface [30-33], presence of various oxides of iron, e.g., FeO, $Fe_3O_4$, $Fe_2O_3$ have been reported using a variety of experimental techniques [2,34-37]. This large discrepancy among various results may partly be attributed to the limited sensitivity of the techniques used in detecting a few monolayer of possible oxide at the buried interface of Fe/MgO. In the present work, we report soft x-ray absorption study of the interface between MgO and Fe studied *in situ* during the deposition of the film itself. Soft x-ray absorption spectroscopy is a highly sensitive technique which can detect even a mono layer of a material. Various oxides of iron can be distinguished very reliable through the spectral shape and chemical shift [38-40]. Ex-situ soft x-ray absorption studies have been reported in the literature to elucidate the Fe/MgO interface [30,36,41]. However it becomes difficult to reliably extract the information about a few monolayer thick Fe/MgO interface in the background of the signal from the bulk of iron layer. In the present study, *in situ* SXAS measurements during deposition of Fe film, allows us to selectively look at the interface and its evolution with film thickness. We have been able to characterise the nature of the interfacial oxide of iron as well as make a quantitative estimate of its thickness. It is found that the interfacial reaction crucially depends upon the state of the MgO surface.

## 2. EXPERIMENTAL DETAILS

Two different substrates were used for the experiment; (i) single crystalline MgO (001) substrate and (ii) a polycrystalline MgO film on Si substrate. MgO film on Si substrate was deposited by ion beam sputtering of an MgO target (99.99% purity) in a sputtering chamber with base pressure of $10^{-7}$ mbar and using a beam of 1000 eV Ar ions produced using a Kaufman type broad beam ion source. The film was characterized using x-ray diffraction and x-ray photoelectron spectroscopy for structure and stoichiometry respectively.

Soft x-ray absorption measurements were done at BL- 01 beamline of Indus 2 synchrotron radiation source, Indore [42]. For doing *in situ* deposition of Fe film in the beamline, a UHV deposition chamber with one inch sputtering source was attached to the beamline with a gate valve in between. The schematic of the chambers with top view is shown in Fig. 1. The substrate was mounted on a rotary motion feed through from the top. It could be rotated to face either the sputtering source for thin film deposition, or the soft x-ray beam for absorption measurements. The base vacuum in the chamber was of the order of $2 \times 10^{-8}$ mbar. Before starting deposition, the MgO substrate in the chamber was annealed at about 100$^{o}$C for 1 h in order to remove any possible adsorbed water at the surface. For the deposition of film, the gate valve between the chamber and the beamline was closed and sputter deposition was made on the substrate using Ar gas and a sputtering power of 5W. After depositing the film for a predetermined time, the gas flow was stopped, and the chamber was brought to a base pressure of $2\times10^{-8}$ mbar, before allowing the soft x-ray beam to enter the chamber by opening the gate valve. For calibrating the film thickness, prior deposition of Fe film was made on a Si substrate in the same chamber for 20 min, and the thickness of the deposited film was measured ex-situ using x-ray reflectivity (XRR). The deposition rate came out to be 0.024 nm/s. SXAS measurements were done in total electron yield mode across the L edge of iron in the energy range 700 eV to 740 eV which spans both $L_3$ as well as $L_2$ edges, however, results are presented only for $L_3$ edge.

## 3. RESULTS

### 3.1 Fe on MgO film

XPS measurements were done in order to determine the stoichiometry of the MgO film (Fig. 2). Before doing measurements, the surface was cleaned by sputtering with 100 eV Ar ions for 5 minutes. The stoichiometry of the film as determined from the relative areas of Fe and O peaks comes out to be $Mg_{55}O_{45}$. Crystallite size as obtained from X-ray diffraction (not shown) using Scherrer formula came out to be 4.1±0.2 nm.

Preliminary study of possible interfacial interaction between Fe and MgO was done using MOKE measurements. For this purpose, a wedge-shaped film of iron was sandwiched between two layers of MgO film, as shown in the inset of Fig. 3. X-ray reflectivity measurements were done at different points along the length of the sample in order to determine the actual thickness gradient of iron layer, it was found to be 0.1 nm/mm. Longitudinal MOKE measurements were done as a function of the thickness of Fe film by moving the laser spot along the length of the sample. Fig. 3 gives the saturation value of MOKE signal as a function of iron layer thickness. For film thickness much less than the penetration depth of the laser beam, strength of the MOKE signal is expected to be a linear function of film thickness given by the equation [43].

$$\phi = \frac{4\pi n_{sub} Q d \theta}{\lambda (1 - n_{sub}^2)}, \qquad (1)$$

$Q$ being the magneto-optical constant, $d$ the thickness of magnetic layer, and $\theta$ the angle of incidence measured from surface normal. A linear fit to the data in Fig. 3 shows that the line cuts the x axis at 0.33±0.15 nm. Since there are two interfaces of Fe with MgO, thickness of magnetic dead layer at one interface comes out to be 0.16 nm.

Fig. 4 gives the soft x-ray absorption spectrum at $L_3$ edge of Fe as a function of Fe film thickness. One can see that the technique is sensitive enough to detect even a film of average thickness of 0.12 nm. One may note that the $L_3$ spectrum consists of 2 peaks at 708 eV and 709.4 eV respectively. While the peak at 708 eV corresponds to bcc Fe, the peak at 709.4 eV may be attributed to possible interfacial oxide. L edge spectrum of Fe and its various oxides have been studied in the literature in detail [38-40], and can be used to identify the oxide phase present at the interface. From the literature, one finds that in $Fe_3O_4$ the main $L_3$ peak shifts by 1.4 eV with respect to metallic iron, while in $Fe_2O_3$ the shift is about 1.7 eV. Thus the peak at 709.4 eV can be identified with that of $Fe_3O_4$. $L_3$ spectrum of $Fe_3O_4$ consists of a main peak corresponding to $Fe^{3+}$ ions at both octahedral + tetrahedral sites and a shoulder at lower energy due to $Fe^{2+}$ ions [40]. One can see that the SXAS spectrum of 0.12 nm thick film matches very well with that of $Fe_3O_4$. Thus one can infer that at the interface with MgO, $Fe_3O_4$ is being formed. One can see that with increasing layer thickness, the peak corresponding to metallic iron increases in intensity at the expense of the peak corresponding to $Fe_3O_4$. The relative contributions of metallic Fe and $Fe_3O_4$ to $L_3$ spectrum of the film of a given thickness was obtained by fitting the spectrum with a linear combination of the spectra corresponding to metallic Fe and $Fe_3O_4$, as shown in Fig. 4. The fractional contribution of metallic iron was determined by normalizing the areas with the total area under the $L_3$ peak. Fig. 5 gives the resultant contribution of metallic iron as a function of total film thickness. The thickness dependence of the fraction of iron can be used to make a quantitative estimate of the thickness of interfacial oxide layer. It may be noted that the absorption coefficient of escaping photoelectrons in the film is quite high and typical escape depth in metals is in the range of a few nm only. Therefore, while correlating the intensity of the absorption peak with layer thicknesses, one should take into account the absorption of photoelectrons in the film itself. If we take the thickness of oxide layer as d and the total thickness of the film as D, the photoelectrons from $Fe_3O_4$ layer will have to travel a distance of (D-d) in the iron layer so as to escape from the surface. Thus, the number of photoelectrons emanating from the oxide layer which are able to escape from the surface is given by:

$$I_{oxide} = I_0 e^{-(D-d)/\lambda_e} \; ; \; I_0 = A*d, \qquad (2)$$

where, $I_0$ is the total number of photoelectrons emitted from a thickness $d$ of the oxide layer, $\lambda_e$ is escape depth of electrons and $A$ is the number of photoelectrons emitted per unit thickness of the film and will depend upon photo absorption cross-section and the number of iron atoms per unit thickness in the illuminated area. Similarly the photoelectrons emitted from iron layer of thickness (D-d) will also undergo self-absorption and the intensity of escaping photoelectrons emanating from the iron layer will be given by:

$$I_{Fe} = A\lambda_e [1 - e^{-(D-d)/\lambda_e}]. \qquad (3)$$

Thus, the fractional intensity of iron peak in the $L_3$ absorption spectrum can be written as:

$$\frac{I_{Fe}}{I_{Fe}+I_{oxide}} = \frac{\lambda_e[1-e^{-(D-d)/\lambda_e}]}{de^{-(D-d)/\lambda_e} + \lambda_e[1-e^{-(D-d)/\lambda_e}]} \; . \quad (4)$$

The attenuation length of photoelectrons in Fe was taken to be 2.1 nm [44]. Thickness dependence of the intensity of iron peak in Fig. 5 was fitted with the above equation by taking the thickness d of the oxide layer as a fitting parameter. Best fit to the data is obtained for d = 0.19 ±0.02 nm. Thus, at the interface 0.19 nm equivalent of Fe forms oxide $Fe_3O_4$. It may be noted that, $Fe_3O_4$ is ferrimagnetic with a magnetic moment of 92 emu/g which is about 42% that of bcc Fe. Thus, this thickness of interfacial oxide is in agreement with the result of MOKE measurement within experimental error.

**3.2 Fe on MgO (001) substrate**

Prior to inserting in the experimental chamber for Fe film deposition, the surface of MgO (001) single crystal was cleaned by repeated sputtering with 500 eV Ar ions and annealing at 700 °C. This resulted in removal of carbon contamination from the surface and also annealing out of structural defects created due to mechanical polishing and sputtering [45].

Fig. 6 gives the soft x-ray absorption spectrum across the $L_3$ edge of iron. The $L_3$ peak occurs at energy of 708 eV. At smaller thickness, a pre-peak is observed at energy of 705 eV, which quickly disappears with increasing layer thickness. Further, position of the main peak does not vary with thickness; however, it exhibits some broadening at higher thicknesses.

It may be noted that the $L_3$ peak of both Fe and FeO lie at the same energy. FeO spectrum can be differentiated from that of Fe through the presence of a pre-peak around 705 eV and slightly narrower main peak as compared to iron metal [38]. In Fig. 6, the soft x-ray absorption spectrum at 0.3 nm thickness of the film closely matches with that of FeO. At higher thicknesses the pre-peak disappears, and the main peak also broadens, and the spectrum matches with that of Fe metal. Thus, present measurements provide a clear evidence of bonding of Fe with oxygen at the interface with MgO to form FeO. This is in contrast to the interfacial reaction of Fe with MgO film, where $Fe_3O_4$ is formed at the interface. It is not possible to get a quantitative estimate of the thickness of FeO at the interface, as the main peaks of FeO and Fe overlap completely. However, a rough estimate of the thickness of oxide layer can be obtained by comparing the relative heights of the pre-peak and the main peak. In the spectrum of 0.3 nm thick film, height of the main peak is 11 times that of pre-peak. For thickness of 0.7 nm, this ratio becomes 24. Thus, the additional thickness of 0.4 nm has not added to the intensity of pre-peak, suggesting that 0.3 nm is the upper limit to the thickness of the interfacial oxide layer.

## 4. DISCUSSION

High sensitivity of SXAS technique with the ability to detect even a monolayer of Fe, combined with *in situ* deposition has enabled us to selectively look at interfacial layers of Fe on MgO. In the cases of both MgO film and single crystalline substrate only few monolayers at the interface hybridize with oxygen. However the oxides formed in the two cases are very different. At the interface with MgO film, Fe forms $Fe_3O_4$, while on MgO (001) interface FeO is formed. The difference may be attributed to the difference in the surfaces presented for Fe deposition in the two cases: (i) in the case of MgO (001) only (001) facet with stoichiometric composition is available, while, (ii) in case of MgO film, because of its polycrystalline nature, various facets as well as disordered grain boundaries will be available. Taking typical width of a grain boundary as 1 nm [46] and the measured value of 4.1 nm for the grain size, the volume fraction of grain boundaries come out to be 10% [46]. Thus grain boundaries are expected to occupy significant surface area. Further, the MgO film is off-stoichiometric with the composition $Mg_{55}O_{45}$.

Heat of formation of various oxides of Fe is 272.0 kJ/mol for FeO, 824.2 kJ/mol for $Fe_2O_3$, and 1118.4 kJ/mol for $Fe_3O_4$ [47]. Thus, among various oxides of Fe, $Fe_3O_4$ has highest heat of formation. This can explain the formation of $Fe_3O_4$ at the interface with MgO film. In the case of iron film on MgO (001) substrate, we get a clear evidence of formation of FeO. In an earlier study on an identically prepared structure, it was found using nuclear forward scattering that an interfacial layer of ~0.4 nm thickness possesses an average hyperfine field of 30.8 T, as against 33 T for bcc iron metal [48,49]. This was taken as an evidence for hybridization of Fe with oxygen in the interfacial region. Present studies show that indeed interfacial hybridization leads to formation of iron mono-oxide of thickness around 0.3 nm. However, it may be noted that, FeO orders antiferromagnetically with Neel ordering temperature of bulk FeO being 198 K [50]. In thin films the Neel temperature is known to get reduced further [51]. Therefore, at room temperature it is expected to be paramagnetic. Thus present results appear to be in contradiction with the results of NFS. However, in the present case an ultrathin layer of FeO is interfaced with magnetic layer of bcc Fe. Studies have shown that Neel temperature of an antiferromagnetic film can be enhanced significantly due to proximity effect of an interfacing magnetic layer [52]. In case of FeO, it was shown that antiferromagnetism of a thin FeO layer can be stabilized at room temperature only when it is interfaced with ferromagnetic Fe on both sides [53]. More recently, study of the magnetism of interfacial FeO has been done using conversion electron Mössbauer spectroscopy [54]; Epitaxial FeO on MgO (001) has been found to be non-magnetic. However, deposition of further atomic layers of iron on top of FeO drastically modified the magnetic properties of FeO. It was shown that almost 50% of the iron atoms acquire a hyperfine field of 37.8 T which is almost 15% higher than that of iron but does not agree with $B_{hf}$ of any of the known oxide of Fe. Present studies agree with this observation; ~0.3 nm thick FeO layer formed at the interface gets partially magnetically-ordered due to proximity effect with the bcc Fe layer on top of it.

Among all the oxides of iron, FeO has the lowest heat of formation and therefore its formation should not be favoured thermodynamically. This suggests that as the iron atoms arrive at the surface of MgO (001), oxygen atoms are not available freely for them to react with. It may be only at a later stage that oxygen atoms diffuse into iron to form FeO. Thus formation of FeO is dictated by kinetic considerations, rather than thermodynamic considerations.

Radically different interface structure between Fe layer on single crystalline MgO (001) surface and on polycrystalline off-stoichiometric MgO surface suggest that the interfacial oxidation of Fe on MgO is sensitive to the surface state of MgO. It may be noted that in the literature a large amount of contradictory results exist about the structure of the interface between Fe and MgO [2,30,33,34-37]. This contradiction may partly be attributed to the ambiguity arising due to the limited sensitivity of the experimental technique used in these studies to characterize the interfacial region which is only a few monolayer thick, and partly to the variation in the surface structure of MgO used in different studies, in terms of surface contamination, stoichiometry, facet, single crystal or polycrystalline nature etc.

## 5. CONCLUSIONS

In conclusion, *in situ* soft x-ray absorption spectroscopy measurement during growth of iron film on MgO has been used to get quantitative information about possible interfacial oxidation of iron. The interfacial reaction sensitively depends upon the facet presented by MgO layer for the deposition, and its stoichiometry. On a polycrystalline MgO surface, 0.19 nm thick Fe layer gets oxidized to form $Fe_3O_4$, which is thermodynamically the most favored oxide phase. In contrast, on a single crystalline MgO (001) surface, about 0.3 nm thick layer of FeO is formed at the interface. Due to proximity effect of the ferromagnetic Fe layer, magnetic ordering in FeO gets stabilized at room temperature. Thermodynamically, FeO being the least favored phase, the reaction seems to be constrained by the limited availability of oxygen atoms. This strong dependence of the interfacial reaction on the surface condition of MgO can also partially explain the contradictory results reported in the literature on the structure of interface between Fe and MgO.

**ACKNOWLEDGMENTS**

This research work was partially supported by Science and Engineering Research Board (Project No. SB/S2/CMP-007/2013). Measurements at BL01 of Indus 2 were done under a collaborative research scheme with UGC-DAE Consortium for Scientific Research, Indore.

**FIGURE CAPTIONS**

Fig. 1. Schematic diagram (top view) of the chamber used for doing *in situ* SXAS measurements during film deposition. The chamber is connected to the beam-line through a UHV gate valve. Magnetron sputtering source was mounted on a port diagonally opposite to the port for x-ray beam. Substrate is mounted on a rotary motion feed through.

Fig. 2. Core level XPS spectra (a) Mg 2s and (b) O 1s of MgO film. The solid lines are fit to the experimental data.

Fig. 3. Intensity of Kerr signal as a function of the thickness of iron film sandwiched between two layers of MgO. The structure of the film used for the measurement is shown in inset.

Fig. 4. $L_3$ absorption spectrum of iron as a function of iron film thickness (D) on polycrystalline MgO film. Each spectrum is fitted with a linear combination of the absorption spectra corresponding to bcc Fe and $Fe_3O_4$.

Fig. 5. Fractional area of metallic iron in the $L_3$ absorption peak as a function of iron film thickness. The continuous curve is the best fit to experimental data using eq. (4).

Fig. 6. $L_3$ absorption spectrum of iron as a function of iron film thickness deposited on MgO (001) substrate. The Inset compares the spectrum of 0.3 nm thick film (-) with that of 4 nm thick film (----).

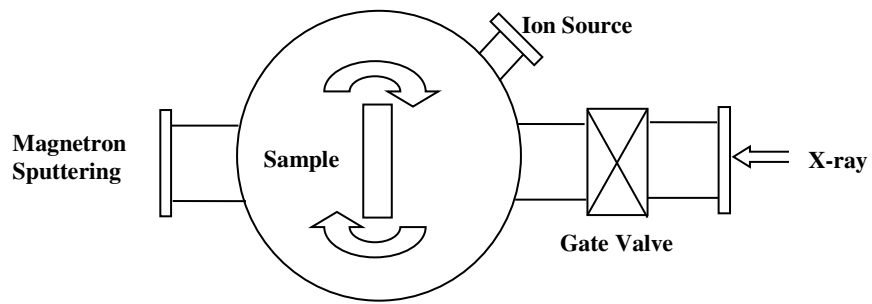

Fig. 1

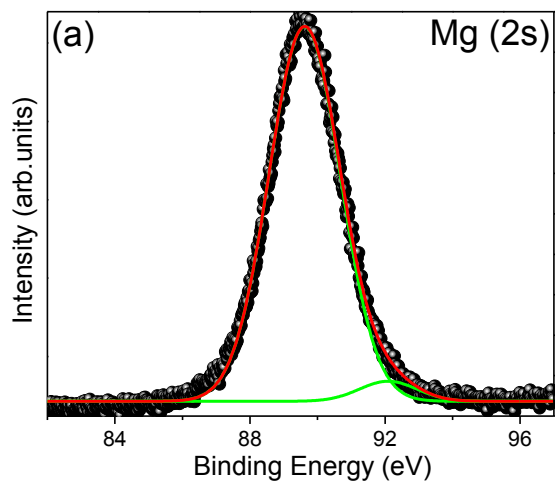 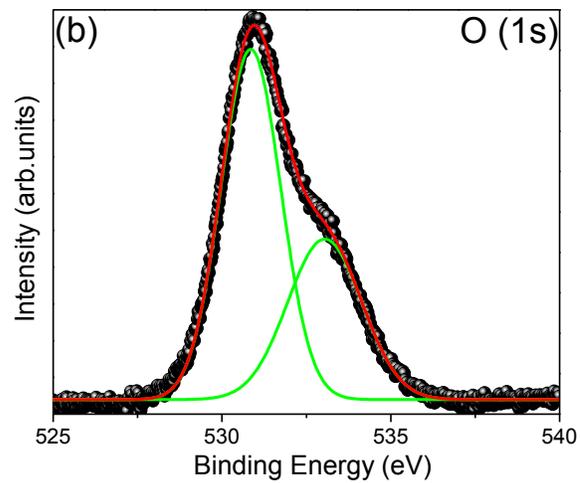

Fig. 2

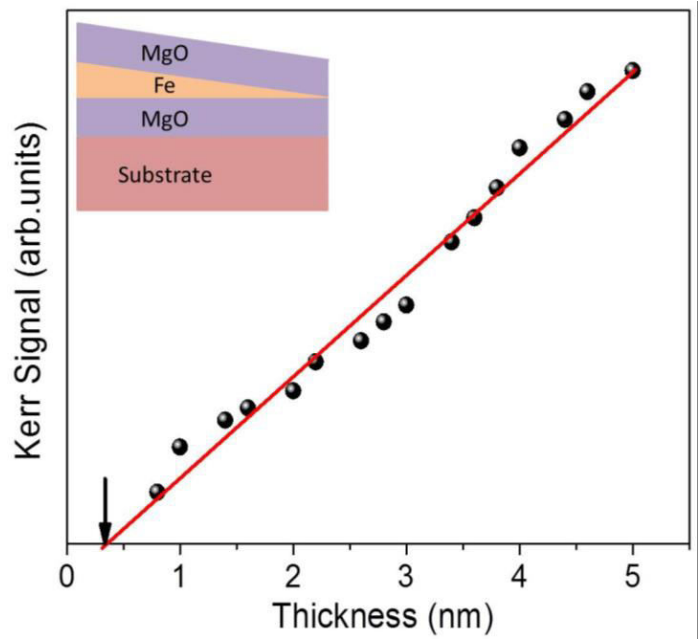

Fig. 3

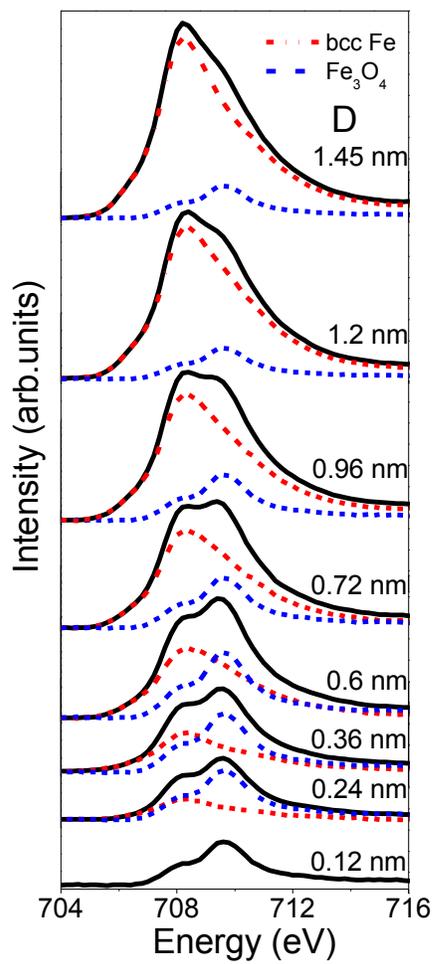

Fig. 4

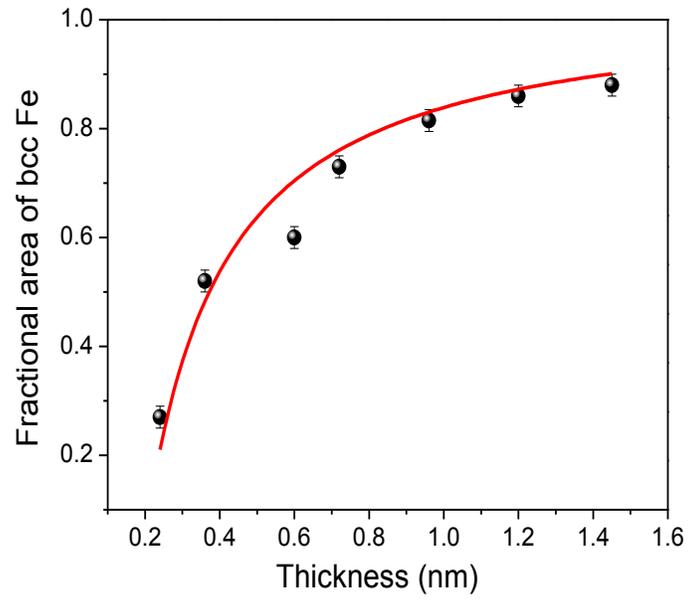

Fig. 5

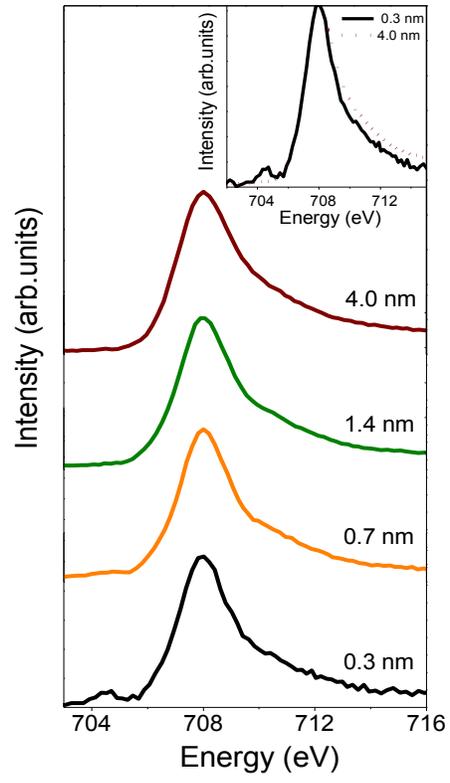

Fig. 6